\documentclass[journal]{IEEEtran}
\usepackage{graphicx}
\usepackage{amsmath}
\usepackage{amssymb}
\usepackage{amsbsy}
\usepackage{multirow}

\begin{document}
%
\title{Traditional Machine Learning for Pitch Detection}
%

\author{Thomas Drugman, Goeric Huybrechts, Viacheslav Klimkov, Alexis Moinet}    

\markboth{IEEE Signal Processing Letters}%
{Shell \MakeLowercase{\textit{et al.}}: Bare Demo of IEEEtran.cls for Journals}

\maketitle

\begin{abstract}
Pitch detection is a fundamental problem in speech processing as F0 is used in a large number of applications. Recent articles have proposed deep learning for robust pitch tracking. In this paper, we consider voicing detection as a classification problem and F0 contour estimation as a regression problem. For both tasks, acoustic features from multiple domains and traditional machine learning methods are used. The discrimination power of existing and proposed features is assessed through mutual information. Multiple supervised and unsupervised approaches are compared. A significant relative reduction of voicing errors over the best baseline is obtained: 20\% with the best clustering method (K-means) and 45\% with a Multi-Layer Perceptron. For F0 contour estimation, the benefits of regression techniques are limited though. We investigate whether those objective gains translate in a parametric synthesis task. Clear perceptual preferences are observed for the proposed approach over two widely-used baselines (RAPT and DIO). 
\end{abstract}


%



\section{Introduction}\label{sec:intro}

Pitch detection (also known as pitch tracking) refers to the task of estimating the contours of the fundamental frequency F0. It is a fundamental problem in speech processing, since pitch information is used in various applications such as text-to-speech, speech recognition, voice assessment, speech perception, speech transformation, language acquisition, speech analysis or speaker identification. 

Dozens of papers on pitch detection are published every year. Good review articles can be found in \cite{Gerhard} and \cite{GlottalReview}. Finding the best acoustic representation has gotten much attention. Features can be extracted in the time \cite{ZCR, Yin, AMDF, RAPT, SIFT, YAPPT, Praat}, spectral \cite{HarmonicProduct, SHR, SRH} or cepstral domain \cite{Noll}, or derived from auditory models \cite{Seneff, AMPEX}. Various other papers have focused on pitch tracking for specific use cases: robustness in noisy environments \cite{SRH, Robust1, Robust2}, real-time constraints \cite{RealTime1, RealTime2}, or particular applications (e.g. singing voice \cite{Singing}, guitar effects \cite{RealTime1} or bird vocalization \cite{Birds}). Finally, there is a recent trend in using deep learning for pitch tracking in noisy environments. The deep classifier is fed either with spectral acoustic features \cite{HanWang, Liu, BLSTM} or directly with the waveform samples \cite{Verma, CREPE}. Classifiers that have been investigated include feed-forward \cite{HanWang, Verma}, recurrent \cite{HanWang, BLSTM} or convolutional \cite{CREPE} neural networks. The classes that are predicted consist of the bins of the discretized log-F0 scale and possibly a state for unvoiced or non-speech frames.

This paper proposes a pitch detection algorithm based on engineered features and traditional Machine Learning (ML). Our goal is to push accuracy as far as possible on clean recordings, as our final use case is high-quality speech synthesis. Nonetheless, we expect that the proposed method would translate well to noisy environments as long as one takes care of a possible mismatch between training and testing conditions. The main contributions of this paper are: \emph{i)} we consider voicing detection as a classification problem and F0 estimation as a regression problem, and apply traditional ML to both tasks; \emph{ii)} we use features from multiple domains, propose to extract them also from the mean-based signal, and quantify their discrimination power; \emph{iii)} we perform a comprehensive experimentation and observe a 45\%  relative reduction of the voicing errors compared to the best baseline, \emph{iv)} we show that very good results can be obtained with little to no supervised data, \emph{v)} we show a significant perceptual improvement after vocoding. 


\section{Method}\label{sec:Method}

Some techniques below are based on clustering and do not require labeled data. Some others are based on supervised learning. In this case, we assume that a training database including Electroglottograph (EGG) recordings is available. 

\subsection{Acoustic Features}\label{ssec:Features}

Our goal is to extract acoustic features in a diversity of domains to exploit their possible complementarity. In the time domain, the Zero-Crossing Rate (ZCR), the peak of the autocorrelation (AC) function and the Clarity feature \cite{Clarity, Clarity2} derived from the AC function are extracted. In the frequency domain, we use the spectral slope between 1 and 7 kHz, as well as the energy-normalized and unnormalized versions of the Summation of the Speech Harmonics (SSH and SSH$^{*}$) which takes the effect of subharmonics into account \cite{SRH}. We also include similar features calculated on the residual signal (SRH and SRH$^{*}$) as proposed in \cite{SRH}. The residue is obtained after extracting the Mel Generalized Cepstral (MGC) coefficients and applying inverse filtering. Finally, the Cepstral Peak Prominence (CPP, \cite{CPP}) is also added.

Besides those 9 features, we propose to extract the same set of descriptors (except SRH and SRH$^{*}$) from the Mean-based Signal (MS). The MS was proposed in the SEDREAMS algorithm \cite{SEDREAMS} for glottal closure instant detection and ensured high identification rates. The calculation of the MS requires a rough estimate of the average F0 for the speaker to determine the window length. When extracting the full set of 16 features, we therefore apply a two-pass scheme in which: \emph{i)} the first 9 features are extracted from the speech signal, \emph{ii)} a rough value of the averaged F0 is estimated from the AC feature, \emph{iii)} the MS and its derived features are calculated. All features are extracted using 30ms-long Hanning windows with 5 ms shift.

\subsection{Voicing Detection}\label{ssec:Voicing}

Voicing detection refers to the task of discriminating whether the signal is voiced (i.e. pseudo-periodic) or not. It is thus a binary classification problem. To achieve that, the acoustic features are fed either to a clustering algorithm or to a supervised classifier.

As clustering approaches, we have tried both K-means and Gaussian Mixture Models (GMMs), forcing the number of clusters to be 2. The voiced cluster can be identified by simply analyzing the feature distribution for both clusters. Those clustering algorithms can be run either directly on the target speaker (T suffix), or on a multi-speaker (M suffix) database. K-means was used to initialize the GMM parameters. Diagonal covariance matrices are used, as we did not observe any gain in using full matrices.

As supervised techniques, a diversity of binary classifiers has been investigated: Logistic Regression, K-Nearest Neighbors (KNN), Adaboost, Random Forest, Support Vector Machine (SVM) and a Multi-Layer Perceptron (MLP). We used a logistic regression regularized with an L2 penalty. For KNN, the 5 nearest neighbors of a point are used to draw the classification decision. Adaboost is a famous boosting algorithm exploiting an ensemble of weak learners on modified versions of the data. Those weak learners are decision trees in our experiments and a maximum of 50 estimators is set before finishing the boosting. Random Forest is also an ensemble technique based on a set of trees built from a bootstrap sampling of the training set. Ten trees have been used in our experiments. For SVM, various kernel functions have been tried: linear, polynomial, radial basis or sigmoid. Not much differences were observed and the conventional radial basis function is used in our experiments. Finally, for the MLP, different network architectures and activation functions have been investigated. We found out that not more than 2 hidden layers with very few units (5 to 20) give good results, and that increasing the number of layers or parameters does not help due to overfitting. Changing the type of activation function only gave marginal differences. The MLP used in our experiments consists of 2 hidden layers of 20 and 10 rectified linear units.

\subsection{F0 Estimation}\label{ssec:Estimation}
In the voiced segments, a F0 value must be estimated for each frame. This can be seen as a regression problem where F0 candidates can be fed as input features. Seven F0 candidates can be used in our workflow and correspond to the peak locations of $F0_{AC}$, $F0_{SSH}$, $F0_{SRH}$, $F0_{CPP}$, $F0_{AC-MS}$, $F0_{SSH-MS}$ and $F0_{CPP-MS}$. On top of those 7 estimators, the 16 voicing measurements presented in Section \ref{ssec:Voicing} can also be used. Four regression methods were tried: linear regression, decision tree, KNN and MLP-based regression (using a L2 loss, hereafter denoted $MLP,reg$). Besides regression approaches, we have experimented with: \emph{i)} a MLP classifier to predict for each frame the index of most accurate F0 estimator (hereafter denoted $MLP,idx$), \emph{ii)} a simple median filter that calculates the median of the F0 estimators.

\section{Experimental Protocol}\label{sec:Protocol}
Our dataset consists of 3 internal and 3 external voices. Our internal voices are from female speakers and correspond to Amazon Alexa's voices in Germany and Japan, as well as a Japanese voice from Amazon Polly. The 3 external voices are the BDL (US male), JMK (Canadian male) and SLT (US female) speakers from the CMU ARCTIC database \cite{ARCTIC}. Respectively 110 and 1150 EEG recordings are available per internal and external voice.

It is widely accepted that EGGs can be used to extract the F0 ground truth \cite{Gerhard, GlottalReview, SWIPE}. For this purpose, we used a technique based on peak detection from the differenced EGG as in \cite{GCI}. It proved to work well after extensive manual validation on multiple voices and, to our experience, was more reliable than the SIGMA algorithm \cite{SIGMA} or than using existing pitch tracking algorithms on the EGG data.

A leave-one-speaker-out cross-validation scheme is applied throughout our experiments of supervised techniques. The data from 5 speakers is thus split into training (90\%) and development (10\%). Testing is carried out on the held-out speaker and this operation is performed 6 times with a speaker-wise rotation. For training, we used a balanced set by limiting the ARCTIC voices to get a similar amount of data as for our internal voices. For testing, the whole data available for the speaker are used though. The results are averaged across speakers, putting an equal weight to all speakers. It should be however emphasized that we did not observe strong inter-speaker variabilities and that the conclusions drawn in this article generalized well on our set of voices.

To measure the accuracy of the pitch trackers, we use 4 standards metrics \cite{Chu, SRH}: the Voicing Decision Error (VDE), the Gross Pitch Error (GPE), the Fine Pitch Error (GPE) and the F0 Frame Error (FFE) rates. VDE is the proportion of frames for which an error of the voicing decision is made. GPE is the proportion of frames where the decisions of both the pitch tracker and the ground truth are voiced, and for which the relative error of $F_0$ is higher than a threshold of $20\%$. FPE is defined as the standard deviation (in \%) of the distribution of the relative error of $F_0$ for which this error is below a threshold of $20\%$. FFE is the proportion of frames for which an error (either according to the GPE or the VDE criterion) is made. FFE can be seen as a single measure for assessing the overall performance of a pitch tracker.

Four pitch detectors are used as baselines: RAPT \cite{RAPT}, SWIPE' \cite{SWIPE}, DIO \cite{DIO} and Harvest \cite{Harvest}. RAPT and SWIPE' have been used widely in the literature and are known to be amongst the best state-of-the-art methods. DIO and Harvest are the pitch trackers used in the WORLD vocoder \cite{WORLD}. Additionally, we use the CREPE \cite{CREPE, CREPE_github} pitch estimator which consists of a convolutional neural network trained on more than 30 hours of vocal and instrumental audio. Note that CREPE does not predict the voicing decisions.

\section{Results}\label{sec:Results}

\subsection{Feature Assessment}\label{ssec:FeatureResults}

\begin{table*}
\centering
\begin{tabular}{| c | c | c | c | c | c | c | c | c | c | c | c | c | c | c | c |}
\hline
ZCR & AC & Clar. & SSH & SSH$^{*}$ & SRH & $SRH^{*}$ & tilt & CPP & ZCR,MS & AC,MS & Clar,MS & SSH,MS & SSH$^{*}$,MS & tilt,MS & CPP,MS\\
\hline
0.13 & 0.55 & 0.16 & \textbf{0.84} & 0.83 & 0.48 & 0.66 & 0.40 & 0.39 & 0.53 & 0.71 & 0.63 & 0.71 & 0.63 & 0.38 & 0.05\\
\hline
\end{tabular}
\caption{Normalized mutual information of the features used for voicing detection.}
\vspace{-.5cm}
\label{tab:Features}
\end{table*}

To quantify the amount of relevant information of each feature for voicing detection, we rely on their Normalized Mutual Information (NMI, \cite{InfoTheory}). NMI is defined as the mutual information between a given feature and the classes, divided by the entropy of the classes. NMI measures the discrimination power of each feature individually but does not reflect their possible redundancy or synergy. If the NMI is 0, the feature is completely irrelevant for voicing detection. If the NMI is 1, the feature allows a perfect voicing classification.



The discrimination power of the features is given in Table \ref{tab:Features}. It can be observed that the SSH features bring the most information (NMI around 0.84). ZCR has been widely used in the past, but is clearly seen as a poor discriminator of voicing (NMI = 0.13). The maximum peak of the autocorrelation, which is the basis of various pitch trackers, does a relatively good job (NMI = 0.55). Overall, spectral features turn out to work best for voicing detection, while the cepstral feature (CPP) gave only moderate results (NMI = 0.39). Features derived from the MS appear to also provide a substantial amount of information. It can be observed that extracting the time-domain features on the MS signal boosts their NMI. The same is not true for spectral and cepstral features.

\subsection{Voicing Detection}\label{ssec:VoicingResults}



\begin{table*}[!ht]
\centering
\begin{tabular}{| c | c | c | c | c | c | c | c | c | c | c | c | c |}
\hline
RAPT & SWIPE' & DIO & K-means,T & K-means,M & GMM,T & GMM,M & Log. Regr. & KNN & Adaboost & Rand. For. & SVM & MLP \\
\hline
5.58 & 5.42 & 6.81 & 4.31 & 4.52 & 6.49 & 5.79 & 3.92 & 4.17 & 4.01 & 3.66 & 3.52 & \textbf{3.11}\\
\hline
\end{tabular}
\caption{Voicing decision error rate (in \%) for the baselines and proposed approaches.} 
\vspace{-.5cm}
\label{tab:Voicing}
\end{table*}

We now investigate the accuracy of the proposed algorithms presented in Section \ref{ssec:Voicing} for voicing detection. Table \ref{tab:Voicing} reports the VDE rate for all techniques. Harvest was found to be particularly poor ($VDE=23.59\%$), although we used the original implementation released by its authors with default settings. This is discussed further in Section \ref{ssec:Synthesis}. Across the state-of-the-art methods, RAPT and SWIPE' provide the best results, with a VDE around 5.5\%.


Through our experiments, we found out that adding the MS features only slightly helped (3\% relative improvement), while it turned out to be important to embed contextual frames. Indeed, stacking the previous and next frames to the current one brought a 15\% relative improvement for both unsupervised and supervised approaches. Extending the context to more frames did not help. The results reported in Table \ref{tab:Voicing} are for the whole set of 16 features with a context of $\pm 1$ frame.

Inspecting the proposed unsupervised approaches, K-means clearly outperforms GMM. Running the K-means clustering on the target speaker (T suffix) or on a multi-speaker (M suffix) dataset does not seem to matter much. Our best unsupervised algorithm (K-means,T) achieves a VDE of 4.31\%, which is already a 20\% relative reduction compared to the best baseline (SWIPE'). It should be emphasized that, as unsupervised approach, no EGG data is required by this technique which makes it simple and scalable.

Moving to supervised algorithms allows to further reduce the VDE. Ultimately, the MLP gives the best results with a VDE of 3.11\%. MLP is respectively followed by SVM and the random forest, as somewhat expected. This is a substantial improvement over the baseline techniques, as the voicing errors are reduced by about 45\% relative over the best existing pitch trackers (RAPT and SWIPE').

\begin{figure}[!ht]
  \centering
  \includegraphics[width=0.48\textwidth]{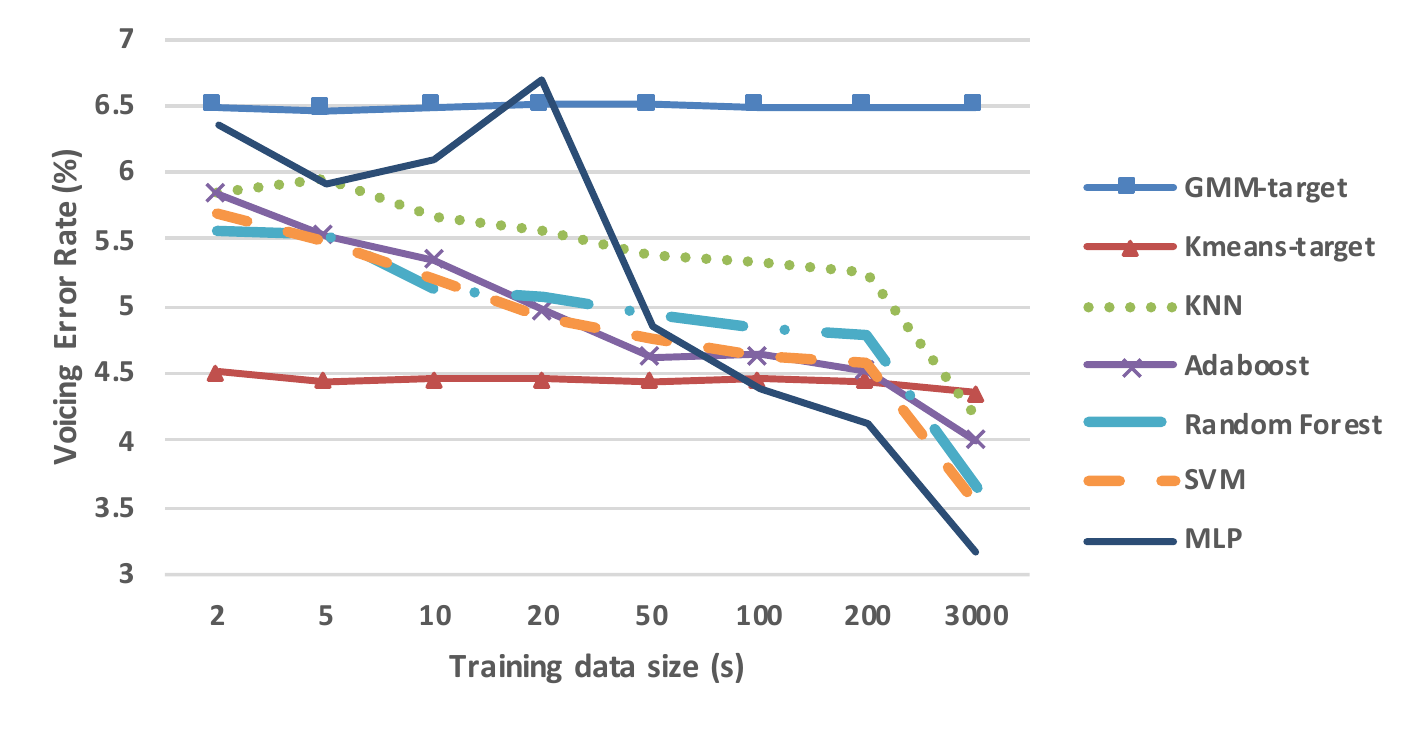}
  \caption{Influence of the amount of training data on the accuracy of the proposed approaches for voicing detection.}
  \label{fig:Datasize}
\end{figure}

The next question we wanted to answer is how much data is needed to train the proposed classifiers. The answer is illustrated in Figure \ref{fig:Datasize}, where the VDE is plotted as a function of the amount of training data (in seconds). To get each data point, we ran 200 experiments by randomly selecting the data and then averaged the results over the 200 outcomes. 

The accuracy of the clustering methods (K-means and GMM) is remarkably flat. Those techniques behave well for limited datasets and increasing the amount of data does not help. It can be interestingly noted that K-means yields the best accuracy up to 100 seconds. For supervised classifiers however, the more data, the better. Outperforming clustering methods is only possible by leveraging more than 5 minutes of EGG data from multiple speakers. If this is not possible, our experiments indicate that one should train a K-means system on the available data.

\subsection{F0 Estimation}\label{ssec:EstimationResults}


\begin{table*}[!htpb]
\centering
\begin{tabular}{| c || c | c | c | c | c || c | c | c | c | c | c |}
\hline
CREPE & $F0_{AC}$ & $F0_{SSH}$ & $F0_{SRH}$ & $F0_{AC-MS}$ & $F0_{SSH-MS}$ & Linear Regr. & Tree & KNN & MLP,reg & MLP,idx & Med. Filt.\\
\hline
3.51 & 2.76 & 3.22 & 6.75 & 3.47 & 16.21 & 3.89 & 3.05 & 2.53 & 3.31 & 2.39 & \textbf{1.95}\\
\hline
\end{tabular}
\caption{Gross pitch error rate (in \%) for the various F0 estimation techniques.}
\vspace{-.5cm}
\label{tab:F0Est}
\end{table*}


Once the voicing segments have been detected using any of the algorithms discussed in Section \ref{ssec:VoicingResults}, we need to estimate the F0 contour within the segment. Out of the 7 pitch estimators generated by our feature extraction scheme, $F0_{CPP}$ and $F0_{CPP,MS}$ were discarded as they led to a GPE rate beyond 30\%. Table \ref{tab:F0Est} reports the GPE for CREPE, the 5 proposed estimators as well as the techniques described in Section \ref{ssec:Estimation}. Across our experiments, we have tried to include the voicing measurements (i.e. the 16 features used for voicing detection) as additional input to the regressor. We have also experimented with reducing the set of F0 estimators, as $F0_{SRH}$ and $F0_{SSH,MS}$ gave higher GPE and could hence impair the final estimate. Finally, we have tried to include contextual values of the estimators to get more reliable predictions. The outcome of those experiments was that: \emph{i)} considering voicing measurements does not help, \emph{ii)} keeping only the 3 best estimators ($F0_{AC}$, $F0_{SSH}$ and $F0_{AC,MS}$) gave better results, \emph{iii)} contextual frames gave more robust results. As a result, we used only the 3 estimators with a context of $\pm 2$ frames.

As seen in Table \ref{tab:F0Est}, only 3 techniques outperform the best single estimator $F0_{AC}$. Those 3 methods are the KNN regressor, the MLP predicting the index of the most accurate F0 estimator and the median filter. It is interestingly observed that a simple combination of the 3 estimators by median filtering outperformed all other approaches (including CREPE). One area that we have not explored though is dynamic programming. Both RAPT and CREPE use Viterbi decoding to smooth the F0 contour. Disabling it for CREPE degraded the GPE from 3.51\% to 4.48\%.

\subsection{Overall Objective Results}\label{ssec:OverallResults}

We summarize now the overall pitch tracking performance along the 4 main objective metrics. For the proposed technique, we used either K-means,T or MLP for voicing detection, followed by the median filter for F0 estimation. Table \ref{tab:Overall} compares the proposed method to the 4 baselines presented in Section \ref{sec:Protocol}. Our proposed algorithm provides a significant reduction of the voicing errors, while maintaining state-of-art GPE and FPE capabilities. It is worth noticing that most pitch trackers avoid estimating F0 when unsure, which leads to better GPE and FPE but worse VDE. Overall, the proposed pitch detector yields a FFE of 3.8\%, which is a 40\% relative reduction over the best baseline (RAPT).

\begin{table}[!ht]
\centering
\begin{tabular}{| c || c | c | c | c |}
\hline
  & VDE & GPE & FPE & FFE\\  
\hline
\hline
RAPT & 5.58 & 1.47 & 3.27 & 6.26\\
\hline
SWIPE' & 5.42 & 1.96 & 3.31 & 6.32\\
\hline
DIO & 6.81 & \textbf{0.63} & \textbf{2.32} & 7.10\\
\hline
Harvest & 23.59 & 1.92 & 2.70 & 24.49\\
\hline
Proposed (K-means,T) & 4.31 & 1.05 & 2.45 & 4.75\\
\hline
Proposed (MLP) & \textbf{3.11} & 1.49 & 2.49 & \textbf{3.80}\\
\hline
\end{tabular}
\caption{Overall accuracy of all pitch detectors.}
\vspace{-.8cm}
\label{tab:Overall}
\end{table}

\subsection{Impact on parametric synthesis}\label{ssec:Synthesis}

The goal of this section is to validate and quantify the impact of the proposed pitch detector in a concrete application. For this, we consider an analysis-modification-synthesis task with the WORLD vocoder using band aperiodicities \cite{D4C}. As modification, we arbitrarily chose to raise the pitch by 20\%. We used 3 possible pitch detectors: RAPT, DIO and the proposed algorithm (using the MLP classifier). Apart from the pitch tracker, all the rest was kept unchanged in the workflow. We have analyzed the effect of the pitch tracker on the final synthesis quality. The systems are compared through subjective preference tests. Twenty participants were involved in each test, each rating 80 pairs of audio samples. These consist of about 5s-long utterances sampled at 24kHz from 8 voices different from those used in training and covering female, male and child speakers. Participants used headphones in standard office conditions.


The results of the preference tests can be found in Table \ref{tab:PrefTest}. The proposed approach turns out to provide clear benefits when compared to the RAPT baseline. The improvement over DIO is also substantial, although to a lesser extent. For both tests, the improvement was statistically significant (p-value $<$ 0.01 with a binomial test). All things except the pitch detector being the same, it is remarkable to observe how the accuracy of a pitch detector can have an impact on the output of a vocoder. Most of the gains stem from proper voicing decisions. It can be wondered why DIO seems to be better than RAPT (although we did not compare them directly) for vocoding, albeit RAPT provided better objective results in Section \ref{ssec:OverallResults}. The explanation lies in the asymmetry of the voicing errors. The 6.8\% error rate of DIO can be broken down into: \emph{a)} 2.1\% of actually voiced frames being detected as unvoiced, \emph{b)} 4.7\% of actually unvoiced frames being detected as voiced. Perceptually speaking, after synthesis, category \emph{a} is more dominant as the effects of a noisy excitation instead of a periodic one can be dramatic. For RAPT, those rates are respetively 3.1\% and 2.5\%, which explains that DIO might be better than RAPT for a vocoding application. Note that the breakdown is extremely skewed for Harvest (0.3\% vs. 23.3\%), which was especially designed for vocoding. The fact that DIO and Harvest generate more voiced segments than other pitch trackers was also mentioned in \cite{Harvest}.

\begin{table}[!ht]
\centering
\begin{tabular}{| c || c | c | c |}
\hline
  & Baseline preferred & Equal & Proposed preferred\\  
\hline
RAPT & 14.8\% & 33.1\% & 52.1\%\\
\hline
DIO & 21.2\% & 45.8\% & 33\%\\
\hline
\end{tabular}
\caption{Results of the preference tests after vocoding.}
\vspace{-.8cm}
\label{tab:PrefTest}
\end{table}

\section{Conclusion}\label{sec:conclu}

While the trend is towards Deep Learning fed with raw audio, this paper showed that an accurate pitch tracker can be achieved with few data using engineered features and traditional ML. We extract acoustic features from multiple domains (time, spectrum and cepstrum) and to feed them to a classification algorithm to predict voicing decisions. The classification can be unsupervised or supervised. Compared to the best state-of-the-art baseline, K-means clustering and a MLP classifier provided respectively a 20\% and 45\% relative reduction of voicing errors. K-means requires as little as a few seconds of speech, while the MLP has a decent accuracy with 1 minute of EGG recordings. In order to predict the F0 contour within the voiced regions, we showed that a simple median filter on the F0 estimators outperforms more complex solutions in terms of gross pitch errors. Finally, we showed that the proposed pitch detector provides also substantial gains in a speech synthesis task.



\newpage

\bibliographystyle{IEEEtran}

\bibliography{mybib}

\end{document}